\documentclass[12pt]{iopart}
\usepackage{hyperref}
\usepackage{graphicx}
\usepackage{bm}

\newcommand{\beq}{\begin{equation}}
\newcommand{\eeq}{\end{equation}}
\newcommand{\beqa}{\begin{eqnarray}}
\newcommand{\eeqa}{\end{eqnarray}}
\newcommand{\matr}[1]{{\rm \mathbf{#1}}}

\newcommand{\ve}[1]{{\bm #1}}

\newcommand{\dd}{{\rm d}}

\begin{document}

\title[Expanded boundary integral method]{Expanded boundary integral method and 
chaotic time-reversal doublets in quantum billiards}

\author{G Veble$^{1,2}$ , T Prosen$^{1}$ and M Robnik$^{2}$}
\address
{$^{1}$ Physics Department, FMF, University of Ljubljana, Jadranska 19, SI-1000 Ljubljana, Slovenia}
\address{
$^2$ CAMTP - Center for Applied Mathematics and Theoretical Physics, University of Maribor, Krekova 2, SI-2000 Maribor, Slovenia}
\ead{\mailto{gregor.veble@fmf.uni-lj.si}, \mailto{tomaz.prosen@fmf.uni-lj.si}, \mailto{robnik@uni-mb.si}}
\date{\today}

\begin{abstract}
We present the expanded boundary integral method for solving the planar Helmholtz problem, which combines the ideas of the 
boundary integral method and the scaling method and is applicable to arbitrary shapes. We apply the method to a 
chaotic billiard with unidirectional transport, where we demonstrate existence of doublets of chaotic eigenstates, 
which are quasi-degenerate due to time-reversal symmetry, and a very particular level spacing distribution that attains a chaotic Shnirelman peak at short energy ranges and exhibits GUE-like statistics for
large energy ranges.  We show that, as a consequence of such particular level statistics or algebraic tunneling 
between disjoint chaotic components connected by time-reversal operation, 
the system exhibits quantum current reversals.
\end{abstract}

\submitto{\NJP}

\maketitle

\section{Introduction}

The solution of the planar Dirichlet problem for the Helmholtz equation has
served as one of the principal numerical setups for verifying and 
demonstrating the main ideas of quantum chaos 
\cite{mcdonelandkaufmann79,Heller84,bgs84,robnik84,bogomolny88,prosenandrobnik93,liandrobnik94}. 
The reason is that this problem, usually referred to as the {\em quantum billiard}, is one of the simplest quantum 
Hamiltonian stationary problems for which a good quality spectra of highly excited eigenstates can be obtained for 
various, even non-integrable geometries.

In literature there exists a good account of numerical techniques to tackle the 
problem \cite{cohen}. 
Apart from many ingenious ideas developed for specific 
geometric setups, there are two competitive general purpose approaches: 
(i) Boundary integral method (reviewed in \cite{Baecker}) or
(ii) Heller's plane wave decomposition method \cite{Heller84}. While the boundary integral 
method is really a general rigorously based method, the plane wave 
decomposition method is a rather heuristic approach which can fail 
in certain important general cases, e.g. of non-convex geometries \cite{Gutkin03}. 
Still, there exists an extremely efficient
implementation of plane wave decomposition method due to Vergini and 
Saraceno \cite{VerginiSaraceno95}, which makes this method an attractive option in spite
of potential mathematical problems. The crucial new
idea of Vergini and Saraceno was to look for a minimum of a boundary norm 
(an integrated norm of the wavefunction or the field amplitude along the 
boundary of a 2-dimensional domain) in terms of a solution of a 
generalized eigenvalue problem, the gain being that in a single 
computational step a number of accurate eigenvalues are obtained in a 
constant proportion to the dimension of the matrices scaling as
${\cal O}(k)$ where $k=2\pi/\lambda$ is a referential wave-number, in 
contrast to the boundary integral method, where a number of matrix computations 
have to be performed in order to locate a single 
eigenvalue. In both cases each matrix computation is of the order ${\cal O}(k^3)$. 

In this paper we propose to use a similar (scaling) idea in conjunction with 
the rigorous boundary integral method. We develop a completely general 
numerical boundary integral technique which produces a constant 
fraction of $k$ eigenvalues per single matrix operation involving
${\cal O}(k^3)$ scalar operations, thus being asymptotically orders of magnitude 
faster than traditional implementations of boundary integral method.

The details of the method, which comprises the first part of the paper, 
are elaborated in Section 2.

The second main idea of the paper is to apply our method in a chaotic 
quantum billiard whose 
phase space structure is not time-reversal invariant, in a sense
that its chaotic phase space components are not mapped onto
themselves upon the time-reversal operation.
Such a situation can appear in generic convex billiards with generic KAM
structure.
Namely we study the billiard in the domain of non-convex 
and non-simply connected shape, the so-called {\em Monza billiard}.
Our choice of the model system also represents a good benchmark for numerical 
methods since it is likely
one of the most difficult types of the billiard shape that one can think of.
The corresponding classical billiard --- being a curved closed corridor with parallel 
walls --- possesses the property of unidirectional motion, namely the classical
phase space separates into two disjoint ergodic and chaotic components of 
clockwise and counter-clockwise motions. The absence of any geometric (point)
symmetries of the model and the fact that the time-reversal symmetry does not 
preserve the invariant ergodic components, has an interesting consequence, 
namely the existence of algebraically (in effective Planck constant $\hbar$) split 
quasi-degenerate pairs of energy levels. The effective Planck's constant of the system is defined as the ratio of the physical Planck's constant to a typical action of a system. In billiard systems, the effective Planck's constant $\hbar\propto 1/k$ where $k$ is the wave-number. The existence of quasi-degenerate pairs has a dramatic effect on energy level statistics,
in particular since the level splitting is of the same order as the mean level
spacing. The effect can be interpreted as a chaotic analogue of the Shnirelman 
peak \cite{Shnirelman, Shnirelman2} known for a long time for systems with non-time reversible 
KAM islands. 

Another and perhaps even more dramatic effect concerns the long range correlations
among levels which behave according to the Gaussian Unitary Ensemble (GUE) of
random matrices in spite of the fact that the system as a whole possesses
time reversal symmetry. This is explained by the fact that even though 
the eigenfunctions of the Hamiltonian are strictly real, the doublets of 
eigenfunctions of the time-reversal operation $T$, namely $T \psi_{\rm L,R}= \psi_{\rm L,R}$, 
which become preserved in the classical limit even though they are not exactly eigenfunctions
of the Hamiltonian, are complex and hence GUE should be used for modeling spectral statistics 
at large energy ranges.\footnote{
For the general discussion of the role of antiunitary symmetries see \cite{robnik1986}.}
This is a new effect, which to our knowledge, has not been predicted
or known before. 

Detailed discussion of the Monza billiard, its spectral statistics and dynamics, is
given in section 3. In section 4 we discuss our main result and conclude.

\section{Expanded boundary integral method}

The scaling method as proposed by Vergini and Saraceno 
\cite{VerginiSaraceno95} allows one to compute a number of states of a
quantum billiard, or any other system described by the Helmholtz equation
with the Dirichlet boundary condition,
lying close to a chosen reference value of the wave-number $k$ without
losing any state in a chosen interval. 
One of its main advantages is that the matrices to be diagonalized
are not of the dimension of the order of $\propto k^2$, as would
be the case when using the 
usual diagonalization techniques, 
but only of the order $\propto k$.
 While this fact makes it one of the most efficient approaches
to solve for eigenvalues and eigenstates of quantum billiard problems,
especially for high values of $k$, experience shows
that its domain of applicability is rather limited. As shown by Gutkin
in \cite{Gutkin03}, the plane wave decomposition method \cite{Heller84}
on which the scaling method is typically based can in general only be applied to
a convex billiard. 
The scaling method does not necessarily involve a plane wave decomposition and a different basis set can be used, which enables the method to solve problems inaccessible by the plane wave decomposition. One such example is the use of Bessel functions of fractional order to solve the mushroom shaped billiard problem \cite{Barnett}. The choice of the basis in such cases must, however, be specifically tailored to the system.

The boundary integral method, see \cite{Baecker} for a review of the method,
has a similar demand
on matrix size as the scaling method and can in principle be applied to
arbitrary shapes as it is based on the Green's theorem that shows how
the wavefunction inside the billiard can be represented with its
normal derivative on the boundary (in the case of Dirichlet boundary 
conditions). From this representation the integral equation for
the normal derivative at the boundary
$u=\frac{\partial}{\partial n}\psi$ follows as \cite{Baecker}
\beq
u(s)=-2\oint_{\partial \Omega} \frac{\partial}{\partial n} G_k 
\left({\ve q}(s),
{\ve q}(s^\prime)\right)~u(s^\prime)~ds^\prime, \label{eq:integral}
\eeq
where $\partial \Omega$ denotes the boundary of the domain, $s$ is the arc length
parametrization of the boundary. The coordinates of the boundary are given as
${\ve q}(s)$ and $\frac{\partial}{\partial n}  
G_k \left({\ve q},{\ve q}^\prime\right)$ is the inward 
normal derivative of the Green's 
function for the unbound problem at wave-number $k$. Following reference
\cite{Baecker} we may discretize equation (\ref{eq:integral}) on a
finite set of points as determined by their arc length parameters $s_i$, yielding
\beq
\left[{\matr A}(k){\ve u}\right]_i=\sum_j A_{ij}(k) u_j=0, \label{eq:matrixeq}
\eeq
\beq
A_{ij}(k)=l_i \left[\delta_{ij}\left(1- \frac{l_i \kappa_i}{2\pi}\right)
+\frac{ik l_j}{2}
H_1^{(1)}\left(k \tau_{ij}\right) \cos \phi_{ij}\right],
\eeq 
where $u_i$ is the value of the normal derivative at the point ${\ve q}(s_i)$,
$\tau_{ij}=\left|{\ve q}(s_i)-{\ve q}(s_j)\right|$ is the distance
between points  ${\ve q}(s_i)$ and ${\ve q}(s_j)$, 
$\cos \phi_{ij}={\ve n}(s_i)\cdot({\ve q}(s_i)-{\ve q}(s_j))/\tau_{ij}$ gives
the angle between the inward boundary normal $\ve n$ 
at the point ${\ve q}(s_i)$ and 
the distance vector
between the two points 
and $\kappa_i$ is the boundary curvature at the point ${\ve q}(s_i)$,
where positive curvature is taken for concave parts of the boundary. The arc-length of the boundary section centered at ${\ve q}(s_i)$ is denoted by $l_i$. The discretization is characterized by the parameter $b=2 \pi N_p / (k L_b)$, where $N_p$ is the number of discretization points and $L_b$ is the total length of the billiard boundary. The parameter $b$ measures the number of discretization points per wavelength.

Equation \ref{eq:matrixeq} only has solutions for those values of $k$ where
the matrix $\matr A$ becomes singular. While in principle any Green's function 
could be used, the choice of the complex Hankel function of the first kind
is necessary in order to avoid spurious solutions that occur because 
equation (\ref{eq:integral}) only then becomes a sufficient condition
for determining the normal derivative.
The usual approach towards solving this
equation is based on solving for zeros of the Fredholm determinant. It 
therefore yields no more than a single level (if at all) per determinant
calculation and is typically plagued with loss of levels and accuracy
when close to a degeneracy. The latter problem has been resolved by B\" acker
\cite{Baecker} by calculating not the determinant directly but the singular value
decomposition (SVD) of the matrix $\matr A$ and following the behaviour of individual 
singular values.

At this point we introduce a similar procedure to the one presented for
the scaling method, namely trying to determine the solution close to a
chosen reference value of the wave-number $k_0$ by varying $k=k_0+\delta k$.
We may write the equation (\ref{eq:matrixeq}) as a Taylor expansion,
\beq
\left[{\matr A}(k_0)+\delta k {\matr A}^\prime(k_0)+\frac{\left(\delta k\right)^2}
{2} {\matr A}^{\prime\prime}(k_0)+\ldots\right]{\ve u}=0, \label{eq:expansion}
\eeq
where ${\matr A}^\prime(k_0)$ and ${\matr A}^{\prime\prime}(k_0)$ are the first and the
second derivative of the matrix ${\matr A}$ 
with respect to $k$ at the point $k_0$, obtained by taking the
derivatives of each matrix element. 
Let us 
expand $\delta k=\epsilon \delta k_0
+\epsilon^2 \delta k_1+\ldots$ and ${\ve u}={\ve u}_0+\epsilon {\ve u}_1+\epsilon^2 {\ve u}_2+\ldots$ in terms of the order of the formal
 perturbation parameter $\epsilon$, whose powers give the order of the terms with respect to the small parameter $\delta k_0$ and should be set to $1$ in the final result. This also shows that the procedure can only be applied to those eigenvectors ${\ve u}$ whose wave-numbers lie close to the reference value $k_0$. By inserting these expressions into the equation (\ref{eq:expansion}) and ordering
the terms with respect to $\epsilon$ we obtain (dropping the implied argument $k_0$)
\beqa
{\matr A}{\ve u}_0+\epsilon \delta k_0 {\matr A}^\prime {\ve u}_0+ \epsilon {\matr A}{\ve u}_1
+\nonumber 
\\
+\epsilon^2 \delta k_0 {\matr A}^\prime {\ve u}_1 
+ \epsilon^2 \delta k_1 {\matr A}^\prime {\ve u}_0
 +\epsilon^2 \frac{\left(\delta k_0\right)^2}
{2} {\matr A}^{\prime\prime}{\ve u}_0+\epsilon^2 {\matr A}{\ve u}_2 + 
{\cal O}\left(\epsilon^3\right)=0.
\label{eq:powers}
\eeqa
We will now show that taking the first two terms of the above equation,
\beq
{\matr A}(k_0){\ve u}_0=\left(-\epsilon \delta k_0\right) {\matr A}^\prime(k_0){\ve u}_0,
\label{eq:main}
\eeq
gives a consistent starting point for solving the perturbative problem. We may do so as there is a degree of freedom involved as to how to define the terms ${\ve u}_0$ and ${\ve u}_1$ in the perturbation series, and the equation (\ref{eq:main}) therefore defines ${\ve u}_0$. 
\footnote{In analogy to the standard problem of (quasi-)degenerate perturbation theory of quantum mechanics, we do not yet know even the starting eigenvectors ${\ve u}_0$ of the problem, and taking only the leading term is not sufficient as they can only be determined by the splitting due to the perturbation.} 
This equation also shows that, while ${\matr A}$ and ${\ve u}_0$ are themselves of the zero order in the $\epsilon$ expansion, their product is of the order ${\cal O}(\epsilon)$. This is due to the fact that we are only slightly shifted in terms of wave-number $k$ from the exact solution of the equation (\ref{eq:matrixeq}), where this product is exactly $0$.

Looking again at all the terms of the equation (\ref{eq:powers}) with up to the linear order in $\epsilon$ and using the equation (\ref{eq:main}),  we obtain
\beq
\epsilon {\matr A}(k_0){\ve u}_1+{\cal O}\left(\epsilon^2\right)=0.
\eeq
As the matrix $A$ is not singular unless we already chose $k_0$ to be the solution of our problem, the above can be satisfied only for ${\ve u}_1=0$. We can therefore write ${\ve u}={\ve u}_0+{\cal O}\left(\epsilon^2\right)$.

We may improve the accuracy of the levels 
by multiplying the equation (\ref{eq:powers}) with the adjoint
of the left eigenvector ${\ve v_0}$, defined as 
\beq
\left[{\matr A}(k_0)\right]^\dagger{\ve v}_0
=\left(-\epsilon \delta k_0\right) \left[{\matr A}^\prime(k_0)\right]^\dagger {\ve v}_0. \label{eq:adjoint}
\eeq
Due to the equations (\ref{eq:main}) and (\ref{eq:adjoint}), the first four
terms are exactly eliminated in pairs. Although ${\matr A}$ and ${\ve v}_0$ are both of the zero order in terms of $\epsilon$, the product ${\ve v}_0^\dagger {\matr A}={\cal O}(\epsilon)$ is of a higher order, as can be seen from equation (\ref{eq:adjoint}). This means that the last term in the equation (\ref{eq:powers}) when multiplied by the left eigenvector also becomes of the order ${\cal O}(\epsilon^3)$. Only two terms remain, giving
\beq
\delta k_1=-\frac{\left(\delta k_0\right)^2}{2} \frac{{\ve v_0}^\dagger
{\matr A}^{\prime\prime}(k_0){\ve u}_0 }{{\ve v_0}^\dagger
{\matr A}^{\prime}(k_0){\ve u}_0} \label{eq:correction}
\eeq
and therefore $k=k_0+\epsilon \delta k_0+\epsilon^2 \delta k_1+{\cal O}
\left(\epsilon^3 \right)$. 

By calculating the equations (\ref{eq:main}) and (\ref{eq:correction})
for various values
of $k_0$, the chosen 
spacing between the values depending on the desired accuracy, we may 
obtain all levels of a system within some chosen interval. As the 
average density of states varies approximately as $\propto k_0$, by taking
the above error estimate into account, the spacing between various reference
wave-numbers $k_0$ at which the individual calculations are performed
must vary as $\Delta k_0\propto k_0^{-(1/3)}$ if the error of calculating an
individual level
is to remain a constant percentage of the mean level spacing. The method itself
is applicable even for calculating the ground state of the system.

For simply connected domains the above procedure is found to be
robust with respect to 
the shape of the boundary, even in the case of arbitrary corners. 
The domains with
holes require an additional step as spurious solutions
 are found in the spectrum. 
These solutions appear because the transpose of the matrix in 
equation (\ref{eq:matrixeq}), which is here used to
 solve the interior Dirichlet problem,
gives the solution to the exterior Neumann problem as well. As the holes 
of the domain 
represent the exterior of our problem and are themselves 
compact, the left eigenvectors
of our spurious solutions are simply the boundary values of the wavefunctions 
of the Neumann 
problem within the holes. 
We may therefore eliminate the spurious solutions by solving 
the Neumann problem for each of the holes by performing
the same computation for each
hole, where the boundary taken is now just the boundary of the hole in question,
and then discard the levels obtained for each hole from the spectrum of the
total problem. 

\section{Monza billiard}

\begin{figure}
\centerline
{
\includegraphics[width=11cm]{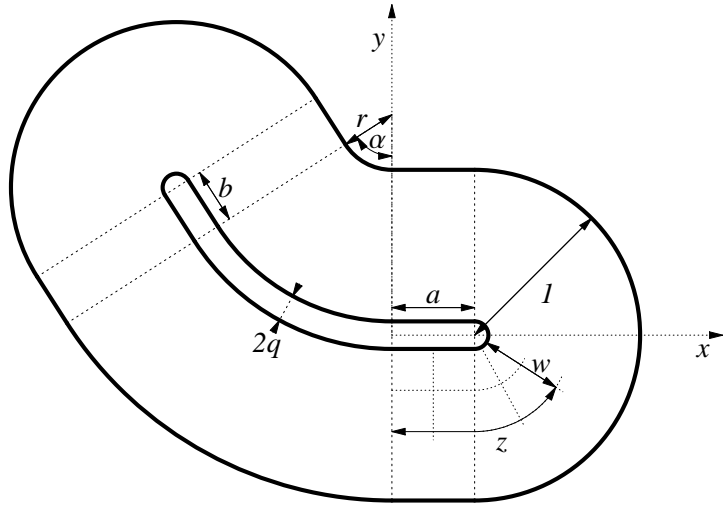}
}
\caption{The definition of the Monza family of billiards. 
The external half-circles
have the radius of $1$, whereas for the internal ones the radius is $q$.
The lengths of the straight segments are $a$ and $b$. The straight segments
are joined by circular arcs with the angle of $\alpha$, and the shortest
of those arcs has the radius $r$. The variables $w\in[0,1]$ and $z\in[0,1)$ (cyclic) as shown are used to parametrize the
billiard. The above shape corresponds to the parameters
chosen in our computation. \label{fig:monza}}
\end{figure}

The method was applied to a billiard that was named the 
Monza billiard due to
its similarity with the famous Italian racetrack. 
It comes from a family 
of unidirectional billiards, as 
defined in \cite{Horvat}. These billiards are shaped as channels with parallel
walls and have the property that the classical trajectories going in one 
direction along the channel cannot reverse this direction. The phase space 
of such systems is
therefore split into two disjoint regions, and the motion within each separate
region can be fully chaotic. The two regions are separated by a one dimensional
family of marginally stable bouncing ball orbits. The shape of the system as
shown in figure \ref{fig:monza} is one of the simplest nontrivial closed 
shapes without any geometric symmetry that belongs to this class of systems. For all the subsequent examples shown, the system parameters
as defined in the figure were chosen to be $q=1/12$, $a=1/2$, $b=1/3$, $r=1/3$, $\alpha=1$ . The billiard was numerically tested to be classically fully chaotic and ergodic within each of the two invariant phase space components \cite{Bunimovich}.

We used the expanded boundary integral method to compute all levels of the Monza billiard up to the wave-number $k=70$. We used the Weyl formula \cite{Weyl} to test that we did not lose any levels in the chosen interval. The discretization parameter was chosen to be $b=12$. The spacing of the reference wave-number $k_0$ between individual runs of the method was chosen to be $\Delta k_0=0.05 k_0^{-1/3}$.

The principle of uniform semiclassical condensation (PUSC, \cite{PUSC,liandrobnik94,robnik98}) states that in the semiclassical limit of effective $\hbar \to 0$ the individual eigenstates of a system correspond to invariant objects in the classical phase space. 
One would perhaps naively expect the eigenstates of a Monza billiard to correspond to either one of the two chaotic components. This, however, can not be the case since states corresponding to either chaotic component would necessarily have a nonzero probability current \cite{robnik1986}. The system, on the other hand, possesses the time reversal symmetry and therefore its non-degenerate eigenfunctions must be fully real and as such can have no probability current.

This seeming contradiction is resolved by noting that in the Monza billiard most eigenenergies are to be found in nearly degenerate doublets. Only the states corresponding to the bouncing ball modes are singlets, with their relative measure vanishing in the semiclassical limit, as they correspond to a classical invariant separatrix between the two chaotic components which itself is the family of bouncing ball orbits corresponding to the particle bouncing perpendicularly between the two walls
and is of measure 0 in the classical phase space. The doublets are a chaotic manifestation of the Shnirelman peak in the level spacing distribution \cite{Shnirelman} that is generally found in all systems with a time reversal symmetry but having no point symmetries. Typically in such systems, the Shnirelman peak is due to the states that correspond to the classical invariant tori which do not map back onto themselves upon the operation of time reversal. There is usually exponentially small tunneling (in terms of effective $\hbar$) between the tori that are connected via the time reversal operation and this is reflected in an exponentially small energy splitting between the states that correspond to the torus pair.
In the Monza billiard, however, the tunneling does not occur between tori but between two chaotic components that are separated by a bouncing ball manifold of measure 0 in phase space. This means that the expected tunneling effects will not be exponentially small but will rather scale as a power law in terms of effective $\hbar$ as the two chaotic components are a distance $0$ apart. In fact, it appears that the average splitting of the pair remains constant as a fraction of the mean level spacing. 
Heuristically, this can be explained by noting that the tunneling amplitude between the states localized in {\em classically invariant} chaotic phase space
components can be estimated in terms of the phase space overlap of the corresponding Wigner functions,
\begin{equation}
V_{\rm LR} = \int \dd^2 {\ve q} ~\dd^2 {\ve p} ~W_{\rm L}({\ve q},{\ve p}) W_{\rm R}({\ve q},{\ve p})
 \propto \hbar \label{eq:wigneroverlap}
\end{equation}
which can be semiclassically estimated to scale as a linear function of an effective Planck constant. 

To demonstrate this, let us fix an arbitrary point in the billiard ${\ve q}$  and locally represent the wavefunction as a combination of random plane waves,
\beq
\psi({\ve x})=\int_0^{\pi}  \dd\varphi ~ f(\varphi) \exp(i {\ve p}_\varphi \cdot ({\ve x}-{\ve q})/\hbar), 
\label{eq:randomwave}
\eeq
where ${\ve p}_\varphi=p_0 (\cos (\varphi+\beta({\ve q})),\sin (\varphi+\beta({\ve q})))$. Due to the unidirectionality, only the waves in one half of the wave-vector space are taken, where $\beta(\ve q)$ denotes the polar angle of the bouncing ball trajectory going through the position ${\ve q}$. We assume a distribution of the stochastic variable $f$ such that
\beq
\left<f^*(\varphi) f(\varphi^\prime)\right>=\rho~\delta(\varphi-\varphi^\prime)\left[\Theta(\varphi)-\Theta(\pi-\varphi)\right], \label{eq:correlator}
\eeq
with
$
\rho=\frac{1}{\pi A} 
$
where $A$ is the area of the billiard and $\Theta$ is the Heaviside step function. Inserting the expression (\ref{eq:randomwave}) into the definition of the Wigner function
\beq
W({\ve q},{\ve p})=
\frac{1}{\left(2 \pi \hbar\right)^2} \int \dd^2{\ve x} ~\psi^*({\ve q}-{\ve x}/2)  \psi({\ve q}+{\ve x}/2)
\exp(-i {\ve p}\cdot {\ve x}/\hbar) 
\eeq
we obtain
\beqa
W_L({\ve q},{\ve p})=\frac{1}{\left(2 \pi \hbar\right)^2} \int_{{\cal D}({\ve q})}  \dd^2{\ve x}\int_0^\pi \dd \varphi
\int_0^\pi \dd \varphi^\prime \nonumber
\\
\cdot  f^*(\varphi) f(\varphi^\prime) 
\exp\left(i \left( ({\ve p}_\varphi+{\ve p}_{\varphi^\prime})/2-{\ve p} \right)\cdot {\ve x}/\hbar \right).
\eeqa
We define the 2D domain
\beq
{\cal D}({\ve q})=\{{\ve x};\ ({\ve q+ {\ve x}/2} \in \Omega) \wedge ({\ve q}-{\ve x}/2 \in \Omega) \}.
\eeq
The expected value of such a Wigner function is
then
\beq
\left<W_L({\ve q},{\ve p})\right>=\frac{1}{\left(2 \pi \hbar\right)^2} \int_{{\cal D}({\ve q})} \dd^2{\ve x}
\int_0^\pi \dd \varphi ~\rho~\exp\left(i \left({\ve p}_\varphi-{\ve p} \right)\cdot {\ve x}/\hbar \right).
\eeq
Performing the integration with respect to ${\ve x}$ the exponential turns into a wide delta function
\beq
\delta_{\sigma} \left({\ve p}\right)=\frac{1}{\left(2 \pi \hbar\right)^2} \int_{{\cal D}({\ve q})} \dd^2{\ve x}
\exp\left(i {\ve p} \cdot {\ve x}/\hbar \right),
\eeq
such that
\beq
\left<W_L({\ve q},{\ve p})\right>=
\int_0^\pi \dd \varphi ~\rho~ \delta_{\sigma} \left({\ve p}_\varphi-{\ve p} \right). \label{eq:wigr}
\eeq
The typical width $\sigma$ of the delta function is nonzero due to the finite size of the system. Its value is
$\sigma\propto \hbar/\ell$ where $\ell$ is some linear dimension of the system independent of $\hbar$. To get an estimate of the overlap in the equation (\ref{eq:wigneroverlap}) we also need to introduce the Wigner function of the state traveling in the opposite direction
\beq
\left<W_R({\ve q},{\ve p})\right>=
\int_\pi^{2 \pi} \dd \varphi ~\rho~ \delta_{\sigma} \left({\ve p}_\varphi-{\ve p} \right)
\eeq
with the integration boundaries complementing those in the expression (\ref{eq:wigr}). If we neglect the correlations we may use the expected values for the Wigner functions in the equation (\ref{eq:wigneroverlap}).
The expected values of the two Wigner functions then overlap due to the width $\sigma$, and the overlap itself is proportional to that width, thus demonstrating the statement (\ref{eq:wigneroverlap}).
\footnote{Note that this estimate is independent of geometrical dimension of the system, i.e. it should be the same for 3D ``tube'' billiards.}

The spectrum can therefore be thought of as a composition of two nearly identical sequences, where their difference remains significant on the scale of the mean 
level spacing. 
It is important to stress that this near degeneracy can not be removed by any means such as desymmetrizing of the billiard as the system chosen does not possess any point symmetries. From this we can deduce that the unidirectional family of billiards will generally exhibit a non-universal spectral statistics. In the level spacing distribution $P(S)$, which is the distribution of spacings $S$ between consecutive levels in an unfolded spectrum where the mean level spacing is equal to one, we expect two main contributions, namely one from the spacings within individual pairs and another from the spacings between consecutive pairs. The first contribution is a widened delta distribution with the weight $1/2$ close to the spacings $S=0$, and the other contribution would be expected to be a stretched GOE (Gaussian Orthogonal Ensemble \cite{Mehta}) contribution. The widening of the second contribution by a factor of two is due to the unfolding procedure which requires that the mean level spacing is equal to one. In a sequence of pairs, the average spacing between centers of different pairs is therefore twice the mean level spacing.

\begin{figure}
\centerline
{
\includegraphics[width=11cm]{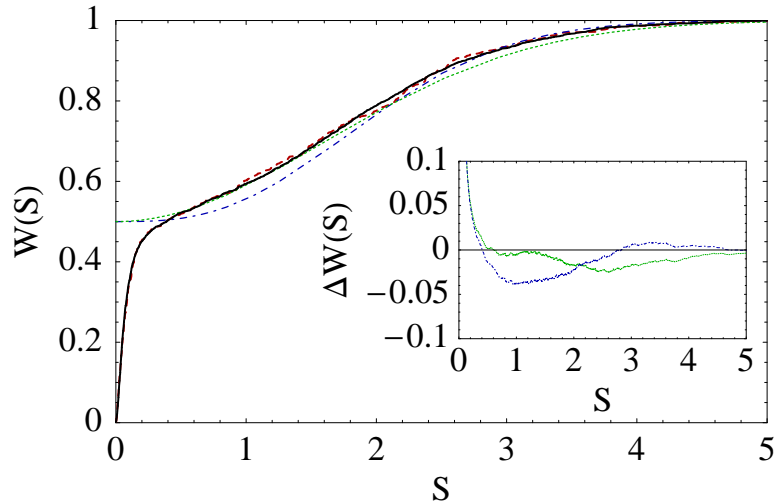}
}
\caption{The integrated level spacing distribution $W(S)$ for the Monza billiard. The numerical results for the $2403$ lowest levels are shown as the full black line. The results for the highest $500$ levels of the same sequence are shown as a red dashed line. The stretched and shifted GOE (dotted green line) and GUE (dot-dashed blue line) predictions are shown as well. In the inset the difference of the GUE and GOE $W(S)$ distributions to the numerical results is shown. \label{fig:ps}}
\end{figure}

The cumulative level spacing distribution
\beq
W(S)=\int_0^S P(S^\prime)\, \dd S^\prime,
\eeq
 for the system with the parameters  $q=1/12$, $a=1/2$, $b=1/3$, $r=1/3$, $\alpha=1$ is given in figure \ref{fig:ps}. The number of levels used in the figure was $N=2403$. For comparison, the same statistics using only the last $500$  levels is shown (dashed red) as well in order to demonstrate that the level spacing distribution indeed does not change with energy (or $\hbar$). Apart from the initial contribution that is mostly due to the spacings within individual pairs, the level spacing distribution then follows the stretched and GOE prediction
\beq
W_{GOE}(S)=\frac{1}{2}\left[1+\int_0^{S/2} P_{GOE}(S') \dd S' \right], \label{eq:WS}
\eeq 
where
\beq
P_{GOE}(S)=\frac{\pi}{2}S \exp\left(-\frac{\pi}{4}S^2\right)
\eeq
is the Wigner surmise \cite{Mehta}. This prediction holds well
for small and intermediate spacings, $S<2$ (shown dotted green), yet at larger spacings the distribution starts to deviate from the GOE prediction but rather approaches the (stretched) GUE prediction (dot-dashed blue). This can be obtained from 
equation (\ref{eq:WS}) but replacing $P_{GOE}$ with
\beq
P_{GUE}(S)=\frac{32}{\pi^2}S^2 \exp\left(-\frac{4}{\pi}S^2\right).
\eeq
While not exactly following this prediction, a clear trend does emerge for the $W(S)$ to move from the GOE to the GUE behaviour at larger spacings.

\begin{figure}
\centerline
{
\includegraphics[width=11cm]{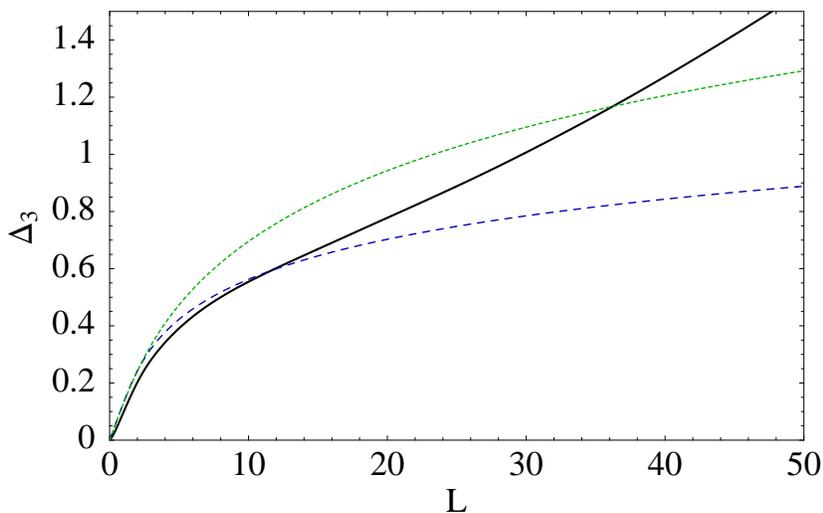}
}
\caption{The $\Delta_3(L)$ spectral rigidity for the lowest 2403 levels of  Monza billiard is shown as a full black line. The scaled GOE (dotted green) and GUE (dashed blue) predictions are given as well. \label{fig:delta}} 
\end{figure}

The longer range spectral statistics are investigated by using the spectral rigidity \cite{Mehta}, which is defined as
\beq
\Delta_3(L)=\left<\min_{A,B}\frac{1}{L} \int_{E}^{E+L} \left[N(x)-(Ax+B)\right]^2\, dx\right>_E,
\eeq
where $N(E)$ is the spectral staircase function whose value increases by one at each (unfolded) energy level. The rigidity measures the average deviations of the spectral staircase function from the best fitting straight line of length $L$. It is given for the spectrum of the Monza billiard in the figure \ref{fig:delta}. We are comparing the numerical results to the $\Delta_{GO(U)E}$ predictions for the random matrix ensembles as defined in \cite{Mehta}. Since most levels come in pairs, the predictions need to be scaled as
\beq
\tilde \Delta_{GO(U)E}(L)=4 \Delta_{GO(U)E}(L/2).
\eeq
The stretching in $L$ is necessary because the average level spacing between level pairs is twice the average level spacing. The factor of $4$ occurs since each step of the spectral staircase function is twice as large for each pair as it would be for an individual level, and the $\Delta_3$ statistics is quadratic in spectral staircase fluctuations.

The results for the $\Delta_3$ statistics of the Monza billiard are given in figure \ref{fig:delta}. At low values of $L$, the results of both predictions as well as the numerical results agree. The numerics undershoots both predictions because the pairs are not exactly degenerate, with the small spacings within the pairs reducing the fluctuations of the spectral staircase function in comparison to the assumed exact degeneracy in the GOE and GUE predictions. At larger distances $L$, the numerics clearly follows the GUE rather than the GOE prediction. At the distances larger than about $L=10$ the numerical $\Delta_3$ statistics starts to increase faster than any of the predictions. This effect could be expected because of the inherent presence of the large bouncing ball mode contribution to the spectrum \cite{bouncing}. Excluding this effect, however, the behaviour of the spectrum at larger distances appears to follow the GUE prediction.

\begin{figure}
\centerline{
\includegraphics[width=5.5cm]{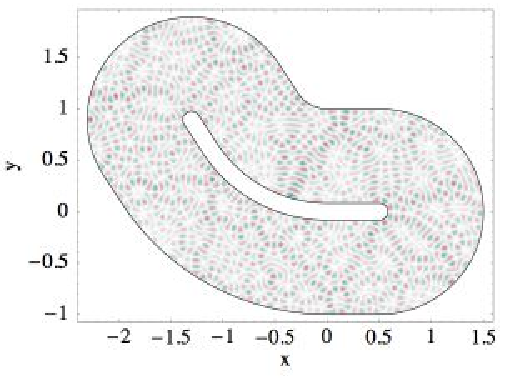}\hspace{0.2cm}
\includegraphics[width=5.5cm]{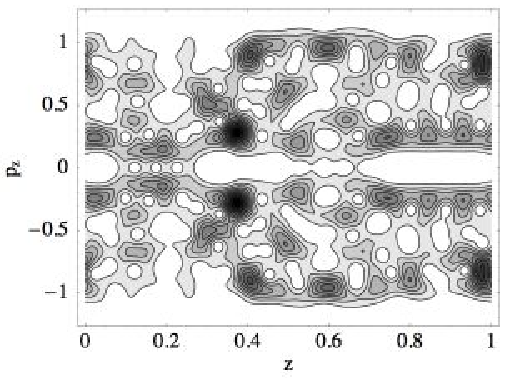}
}
\centerline{
\includegraphics[width=5.5cm]{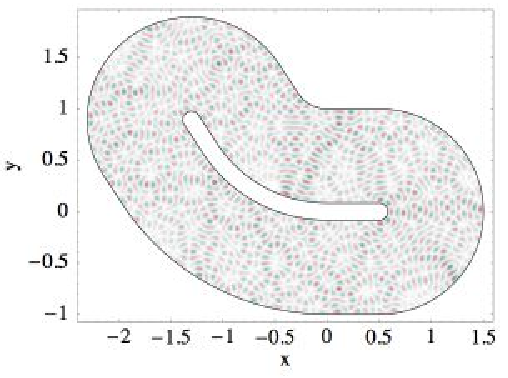}\hspace{0.2cm}
\includegraphics[width=5.5cm]{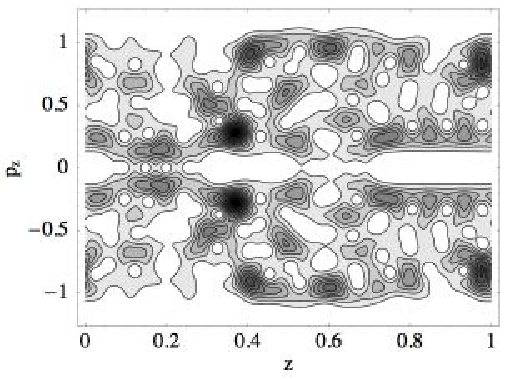}
}
\centerline{
\includegraphics[width=5.5cm]{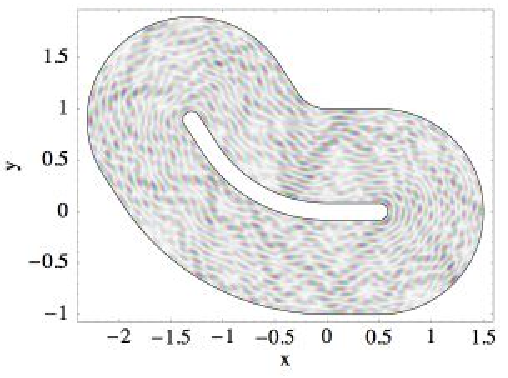}\hspace{0.2cm}
\includegraphics[width=5.5cm]{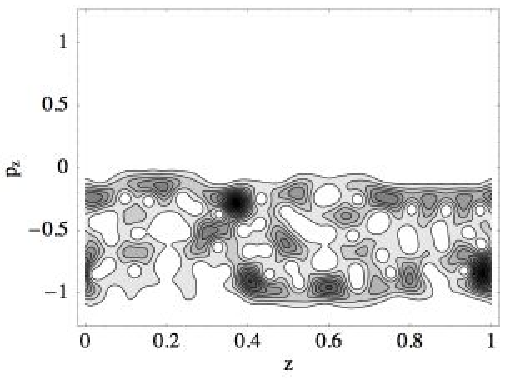}
}
\caption{States corresponding to the nearby pair of states with $k_a=60.0725$ (top) and $k_b=60.0740$ (middle), along with their complex linear superposition (bottom). The Husimi distributions for the wavefunction cross-section along the middle line of the Monza billiard ($w=1/2$, $z\in[0,1)$, see figure \ref{fig:monza}) are shown to the right of each corresponding wavefunction. The darkness of the wavefunction corresponds to the probability density, whereas the hue going circularly from red to green to blue to red again represents the change of the phase of the wavefunction. In the Husimi plots there are 10 contours of constant value shown, starting from $0$ to the maximum value, where on the $x$ axis the coordinate that runs along the length of the Monza billiard $z$ is given, while on the $y$ axis the momentum along this direction is shown in the units of the maximum attainable momentum at the corresponding classical energy.
\label{fig:states}}
\end{figure}

To understand such behaviour we need to look into the properties of state pairs.
In figure \ref{fig:states} we show two states corresponding to a neighbouring pair of states $\psi_a$ and
$\psi_b$. Both states are of course fully real, with the phase switching between $0$ (shown red) and $\pi$ (shown cyan), but a linear combination 
$\psi_{\rm L,R} = (\psi_a\pm i \psi_b)/\sqrt{2}$ of two real states can be shown to have the largest current amplitude. In our case, the direction of the phase change, which itself corresponds to the direction of the current probability, shows that this current flow is clearly seen to be predominantly in a single direction. The complex conjugate of the function is orthogonal to the function itself and corresponds to a current in the opposite direction. The current behaviour is even more clearly exposed by looking at the Husimi plots of the 1D wavefunction cross section along the coordinate $z$ taken at $w=1/2$ (see figure \ref{fig:monza} for coordinate definitions). While for the two eigenstates the Husimi plots are nearly identical, examining the Husimi plot for the complex combination of the two states clearly yields a state that corresponds to a current in a single direction. This is typical for most pairs of states and, while PUSC can not be strictly applied to individual eigenstates, taking a nearly degenerate pair and performing the above complex rotation in this subspace yields states that support the PUSC conjecture.\footnote{A similar analysis applies to (almost) degenerate tori in a KAM-like scenario.}

\begin{figure}
\centerline{
\includegraphics[width=11cm]{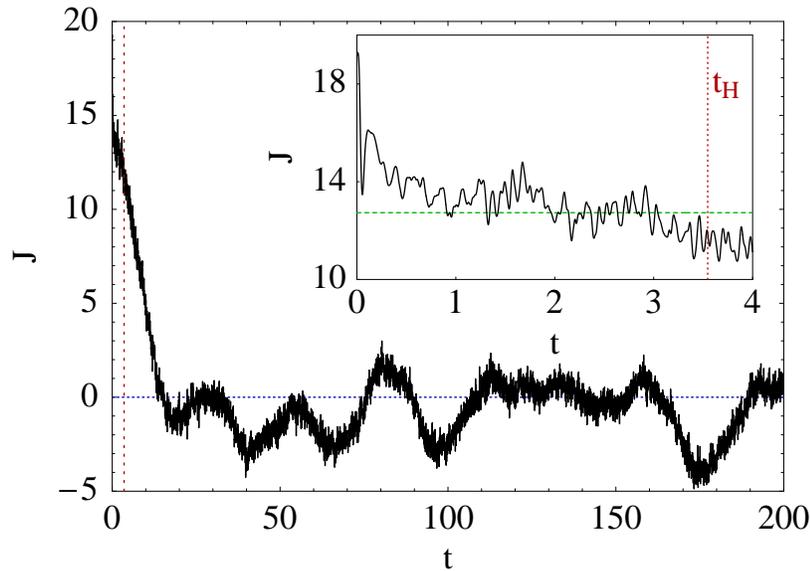}
}
\caption{The expected value of the current operator as a function of time is shown as the full black line. The inset shows the same plot magnified for short times. The dotted blue line shows the value $0$ while the dashed green line gives the classically expected current for the whole chaotic component. The vertical dotted red line indicates the Heisenberg time $t_H=\frac{2 \pi}{\Delta E}$ where $\Delta E$ is the
mean level spacing. \label{fig:current}}
\end{figure}

The nature of the spectrum naturally poses the question of the dynamics of such a quantum system. Classically, a particle traversing the billiard in one direction will keep that direction permanently. Starting with a quantum state such as shown in the bottom part of the figure \ref{fig:states}, as this state is not an eigenstate of the system, it will exactly reverse its character every $t=\pi\hbar/\Delta E$ where $\Delta E$ is the energy splitting of the pair of the two eigenstates forming the initial state. Let us now define the operator for the current along the billiard direction as
\beq
\hat J=-2i \vec e_z \cdot \nabla,
\eeq
where $\vec e_z$ is the unit vector in the direction of the $z$ coordinate in the billiard (see figure \ref{fig:monza}). The initial state is chosen to be a Gaussian wave-packet
\beq
\psi(\vec r,t=0)=\frac{1}{2 \pi \sigma} \exp\left(-\frac{(\vec r-\vec r_0)^2}{2 \sigma^2}+i \frac{\vec p_0\cdot \vec r}{\hbar}\right)
\eeq
with $\vec p_0=(10,0)$, $\vec r_0=(0,1/2)$ and $\sigma=1/5$ with the coordinate frame as shown in figure \ref{fig:states}. We expand this initial state in the basis of all eigenstates up to the wave-vector $k=20$, where these states were obtained using the boundary discretization parameter $b=18$. The evolution equation is taken to be the dimensionless Schr\"odinger equation
\beq
i \partial_t \psi=-\Delta \psi.
\eeq
The expected value of the current operator $J$ as a function of time is shown in figure \ref{fig:current}. As can be seen, for a very short time the initial wave-packet has a large current that up to until the time $t=1$ relaxes to the value $v_0 2/\pi$ (shown dashed green), where $v_0=20$ is the classically expected velocity for the wave-packet with the chosen average energy. Classically, the current is expected to remain at this plateau, but the quantum system experiences a decrease of current that actually turns to fluctuating reversals of the current as it does not stabilize at $0$. This behaviour remains even for times much larger than the ones shown in the figure. The fluctuations are seen to be a fairly large proportion of the initially chosen current.

\begin{figure}
\centerline{
\includegraphics[width=5.5cm]{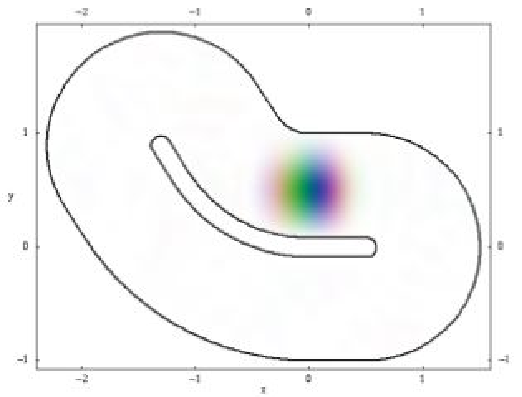}\hspace{0.2cm}
\includegraphics[width=5.5cm]{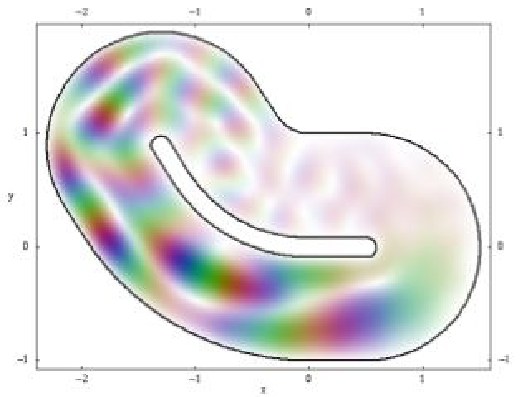}
}
\centerline{
\includegraphics[width=5.5cm]{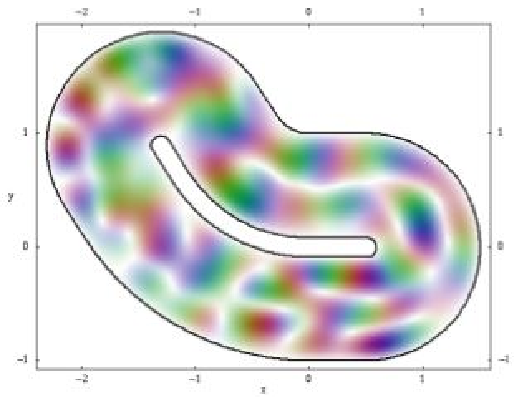}\hspace{0.2cm}
\includegraphics[width=5.5cm]{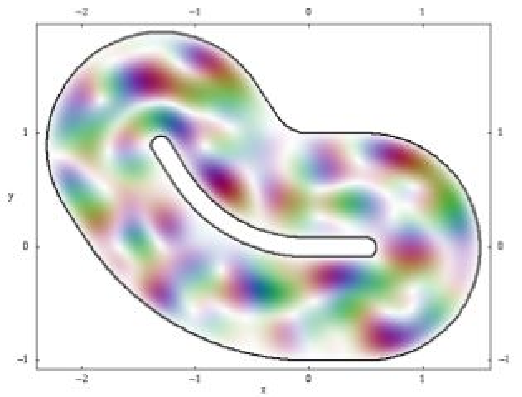}
}
\caption{The snapshots of the video (\href{run:mov6.gif}{GIF animation, 8.0MB}) showing the wave-packet evolution (defined in text) for the times $t=0$ (top left), $1/4$ (top right), $1/2$ (bottom left), $3/4$ (bottom right). The representation is the same as in figure \ref{fig:states}. \label{fig:time}}
\end{figure}

\begin{figure}
\centerline{
\includegraphics[width=5.5cm]{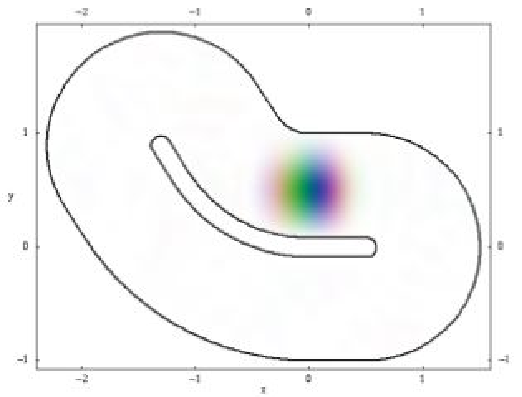}\hspace{0.2cm}
\includegraphics[width=5.5cm]{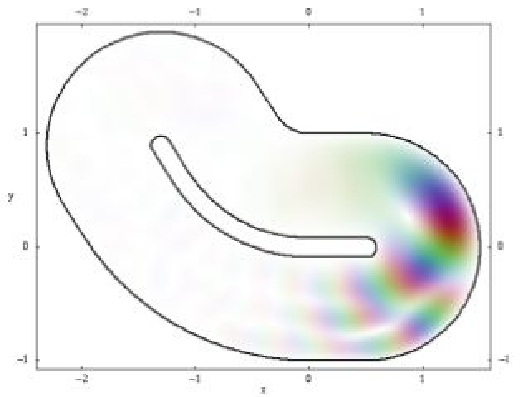}
}
\centerline{
\includegraphics[width=5.5cm]{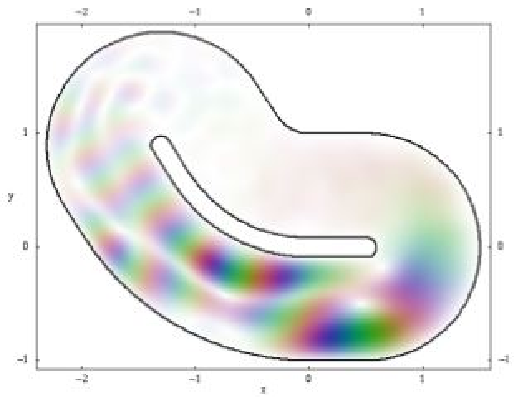}\hspace{0.2cm}
\includegraphics[width=5.5cm]{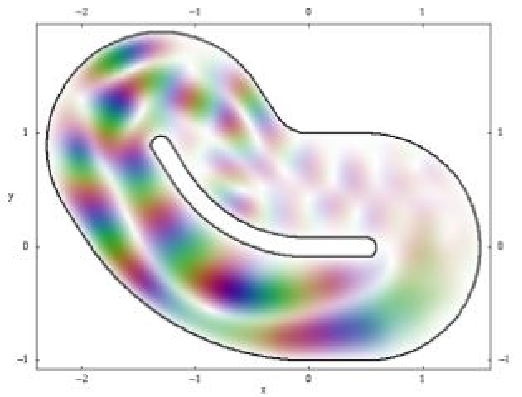}
}
\caption{The snapshots of the video (\href{run:mov7.gif}{GIF animation, 6.2MB}) showing the wave-packet evolution (defined in text) for the times $t=0$ (top left), $0.09$ (top right), $0.18$ (bottom left), $0.27$ (bottom right). The representation is the same as in figure \ref{fig:states}. \label{fig:timeslow}}
\end{figure}

\begin{figure}
\centerline{
\includegraphics[width=5.5cm]{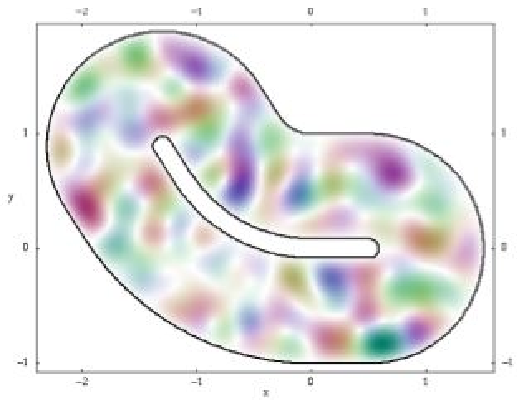}
}
\caption{The snapshot of the video (\href{run:mov8.gif}{GIF animation, 8.9MB}) showing the wave-packet evolution (defined in text) for the time $t=175$. The representation is the same as in figure \ref{fig:states}. \label{fig:timerev}}
\end{figure}

The stills from the video showing the behaviour of the wavefunction for the times up to $t=1$ are shown
in the figure \ref{fig:time}. A slowed down representation for the times up to $t=0.3$ is shown in the figure
\ref{fig:timeslow}. These show the evolution of the wave-packet as it spreads out up to the point where it almost uniformly covers the billiard and therefore corresponds to the wave-packet as spread over the whole unidirectional chaotic component. The phase of the wavefunction clearly indicates that the direction of the quantum current is indeed directed in a single direction along the billiard for all the times shown. In figure \ref{fig:timerev} we show a still from the video representing the evolution of the wavefunction at the time $t=175$ until the time $t=175.3$, where the current as shown in the figure \ref{fig:current} experiences its strongest reversal. From the phase representation the direction of the current is not easily read. The figure \ref{fig:current}, however, shows that the global unidirectional current contribution is quite sizable but the local fluctuations seem to obscure it.

\section{Conclusions}

The goal of the present paper was twofold: 

(i) Presentation of a novel efficient
method for numerical solution of the 2D Helmholtz equation with Dirichlet
boundary condition (e.g. a quantum billiard) which is obtained by combining
 rigorously founded boundary integral method with the ideas
of Vergini and Saraceno \cite{VerginiSaraceno95} of expanding the boundary norm as a function of energy. 

(ii) Application of the new technique to compute the 
energy spectra, their statistical properties, and time evolution in a quantum 
billiard with a chaotic classical limit whose phase space structure is not 
time reversal invariant. Here we mean that the system itself is (globally) time-reversal invariant but
different disjoint chaotic components are not mapped onto themselves
upon time reversal operation. Such situation is possible in the so-called Monza 
billiard due to unidirectionality of the classical motion \cite{Horvat}. We show the presence
of energy level statistics converging to a well defined distribution in the semiclassical limit, 
which is non-universal and exhibits a Shirelman-type peak at small energy ranges and GUE fluctuations at large
energy ranges. We have also computed the time evolution of the
expectation value of the tangential momentum (or particle current) which 
exhibits interesting irregular oscillations due to {\em algebraic} quantum 
tunneling phenomena. The time-scale of the current reversal and consequent oscillations is a constant factor
(typically much larger than unity) of the Heisenberg time. The results have been qualitatively explained in terms 
of a heuristic semiclassical picture, namely the principle of uniform semiclassical condensation combined with 
the nontrivial effects due to time reversal symmetry. 

The phenomenon of GUE-like fluctuations in a time-reversal invariant system, which we discovered at large energy ranges, is
similar to the behaviour observed by Leyvraz, Schmit and Seligman \cite{leyvraz} in systems with point
symmetries lacking real representations. Indeed the two phenomena can be understood to have
some formal similarities. In future explorations it would be interesting also to consider systems which
live in both situations simultaneously, having classically unidirectional motion and point symmetry with 
complex representations.

As the quantum billiard models can nowadays be easily realized
in various experimental setups (e.g. quantum dots, microwave cavities,
optics etc) we expect that predicted effects should be experimentally 
observable.

\ack
We acknowledge useful discussion with A. B\" acker and T.H. Seligman and 
financial support by the Slovenian Research Agency (ARRS), programme P1-0044 and 
grant J1-7347.

\appendix
\section{Merging level sequences}
To obtain the whole spectrum from the individual runs at various reference values $k_0$ a 
method must be chosen to merge all the individual level sequences into a single list. While
in most applications a simple trimming of individual sequences so that they fit with their neighbours
on the energy scale is sufficient, there always exists a possibility that, if a level is very close
to the trimming boundary of a subsequence, it may be accidentally removed, or doubled if, due to the 
numerical inaccuracy, it happens to be present just barely within the neighbouring list as well.
To merge two subsequences $\left\{k_i\right\}$ and $\left\{k_j^\prime \right\}$, 
the first calculated at a reference value $k_0$ and the second at
$k_0+\Delta k_0$, we first define a weight function $\rho(x)$, where $x(k)=\Delta k_0^{-1} (k-k_0)$. 
The function is defined such that $\rho(x\le0)=\rho(x\ge1)=0$, $\rho(1/2)=1$ and that its derivative is bounded such
that $\left| \rho^\prime \epsilon_k \Delta k_0 \right| \ll 1$, where $\epsilon_k$ is the typical numerical error of
an individual level.
We used the function $\rho(x)=4x(1-x)$. First we calculate
\beq
\sigma=\frac{1}{2}\left[\sum_{\left\{k_i\right\}} \rho\left(x(k_i)\right)+
\sum_{\left\{k_j^\prime \right\}} \rho\left(x(k_j^\prime)\right)\right]
\eeq
which gives the estimate of the (weighted) number of levels in the interval 
$\left(k_0,k_0+\Delta k_0 \right)$. If this weighed estimate is smaller than, say $1/2$, we know there are
no levels in the neighbourhood of the middle point of the interval and we may just merge the two 
subsequences by trimming them in the middle and joining them. Otherwise we trim the first sequence by discarding all the levels that are higher than the middle point of the interval and then vary the trimming point $k_c$ for the second sequence in such a way that
\beq
\sigma^\prime(k_c)=\left[\sum_{\left\{k_i<k_0+\frac{\Delta k_0}{2}\right\}} \rho\left(x(k_i)\right)+
\sum_{\left\{k_j^\prime>k_c \right\}} \rho\left(x(k_j^\prime)\right)\right]
\eeq
differs as little as possible from $\sigma$. One should also always demand that $k_c>k_0$ (preferably
even $k_c>k_0+\frac{\Delta k_0}{4}$ to avoid small contributions to $\sigma^\prime$ from the edges of the interval), otherwise possible levels outside the chosen interval may be inadvertently added.



\section*{References}

\begin{thebibliography}{99}

\bibitem{mcdonelandkaufmann79} McDonald S W, Kaufman A N 1979 {\it Phys. Rev. Lett.} {\bf 42} 1189

\bibitem{Heller84} Heller E J 1984 {\it Phys. Rev. Lett.} {\bf 53} 1515 

\bibitem{bgs84} Bohigas O, Giannoni M.-J. and Schmit C 1984 {\it Phys. Rev. Lett.} {\bf 52} 1

\bibitem{robnik84} Robnik M 1984 {\it J. Phys. A: Math. Gen.} {\bf 17} 1049

\bibitem{bogomolny88} Bogomolny E B 1988 {\it Physica D} {\bf 31} 169

\bibitem{prosenandrobnik93} Prosen T and Robnik M 1993  {\it J. Phys. A: Math. Gen.}  {\bf 26} 5365

\bibitem{liandrobnik94} Li B and Robnik M 1994 {\it J. Phys. A: Math. Gen.}  {\bf 27} 5509

\bibitem{cohen} Cohen D, Lepore N and Heller E J 2004 {\it J. Phys. A: Math. Gen.} {\bf 37} 2139

\bibitem{Baecker} B\" acker A 2003 {\it Lecture Notes in Physics} {\bf 618} 91

\bibitem{Gutkin03} Gutkin B 2003 {\it J. Phys. A: Math. Gen. } {\bf 36} 8603

\bibitem{Barnett} Barnett A H, Betcke T 2006 {\it Quantum mushroom billiards} preprint
{\tt nlin.CD/0611059}

\bibitem{VerginiSaraceno95} Vergini E and Saraceno M 1995 {\it 
Phys. Rev. E} {\bf 52} 2204

\bibitem{Shnirelman} Shnirelman A I 1975 {\it Usp. Mat. Nauk} {\bf 30} 265

\bibitem{Shnirelman2} Chirikov B V and Shepelyansky D L 1995 {\it Phys. Rev. Lett.} {\bf 74} 518

\bibitem{robnik1986} Robnik M 1986 {\it Lecture Notes in Physics} {\bf 263} 120

\bibitem{Horvat} Horvat M and Prosen T 2004 {\it J. Phys. A: Math. Gen.} {\bf 37} 3133;
Horvat M, PhD Thesis, University of Ljubljana 2006

\bibitem{Bunimovich} It was suggested by Leonid Bunimovich in a private 
communication that unidirectional billiards with sufficiently long 
straight segments might be provably ergodic within each of the two chaotic 
components

\bibitem{Weyl} Baltes H P and Hilf E R 1976 {\it Spectra of Finite Systems} Bibliographisches Institut Wissenschaftsverlag, Mannheim

\bibitem{PUSC} Berry M V and Robnik M 1984 {\it J. Phys. A: Math. Gen.} {\bf 17} 2413

\bibitem{robnik98} Robnik M 1998 {\it Nonlinear Phenomena and Complex Systems (Minsk)} {\bf 1} 1

\bibitem{Mehta} Mehta M 1991 {\it Random Matrices} Academic Press, New York

\bibitem{bouncing} B\" acker A, Schubert R and Stifter P 1997 {\it J. Phys. A: Math. Gen.}  {\bf 30} 6783

\bibitem{leyvraz} F. Leyvraz, C. Schmit and T. H. Seligman 1996 {\it J. Phys. A: Math. Gen.} {\bf 29} L575

\end{thebibliography}
\end{document}